\documentstyle[preprint,prd,aps,epsf]{revtex}
\begin{document}
\draft
\preprint{July 25, 1998}

\title{Acoustoconductance of quantum contacts}
\author{M. P. Blencowe$^1$ and A. Y. Shik$^{2,3}$}
\address{{}$^1$The Blackett Laboratory, Imperial College, London SW7~2BZ, UK.}
\address{{}$^2$A. F. Ioffe Physical-Technical Institute, 194021 St.
Petersburg, Russia.}
\address{{}$^3$Present address: Energenius Centre for Advanced 
Nanotechnology, University of Toronto, Toronto M5S~3E4, Canada.}

\maketitle
\begin{abstract}

We describe theoretically the  acoustoconductance (AC) of quantum contacts. One 
characteristic of a contact which distinguishes it from a long, uniform
wire is a strong, energy-dependent transmission probability. This has several
consequences for AC. Electrons which are {\it forward} scattered
by phonons can contribute to AC and, furthermore, AC can have positive sign
(i.e. a conductance-increase under the influence of phonons). 
By contrast, for uniform wires only backscattered electrons contribute to AC
which always has negative sign.
\vskip 2cm
\noindent Author responsible for further correspondence: 
M. P. Blencowe, The Blackett Lab., Imperial
College, London SW7 2BZ, England. Fax: +44-171-594-7604. E-mail:
m.blencowe@ic.ac.uk   
\end{abstract}
\vskip 2cm

Non-equilibrium ballistic phonons are a powerful tool for investigating 
the electron-phonon interaction and other electronic properties of 
low-dimensional systems (see, e.g.,  Ref.\ \cite{CKR90}). 
Recently, the first experiments on acoustoconductivity (AC) in quantum point
contacts were performed \cite{naylor} (see also the contribution by A. J. Kent
in this volume).  In this note, we present some of the theory for the 
AC properties of a quantum contact (QC).

We  model the contact as a  wire of length $L$ containing in the middle 
a single narrow barrier with energy dependent transmission probability 
$T_{M k}$, where $M$ denotes the subband label and $k$ is the electon 
wave vector. The total electron energy is $E_{M k}=E_M + \hbar^2 k^2 /2m$. 
This model, which was
considered in Ref.\ \cite{GPF95} for thermal phonons, allows on the one
hand an exact semiclassical (i.e. Boltzmann kinetic equation) solution, 
while on the other hand has in 
common with a QC the main property which distinguishes it from a uniform wire, 
namely a strong, energy-dependent transmission coefficient (see also 
Ref.\ \cite{tot} for a closely related semiclassical treatment of the AC of
nonuniform quantum channels).    

Generalizing the semiclassical calculations of Ref.\ \cite{GPF95} 
to allow for arbitrary 
phonon distribution functions, we obtain the following
phonon-induced correction to the conductance:
\begin{eqnarray}
\Delta G &=& -\frac{2\pi e^2 L}{\hbar k_{\rm B}T}\sum_{MM'}\sum_{{\bf q}, k'}
\int_0^\infty\frac{dk}{2\pi}\ \delta_{q_x,k+k'}\
\delta(E_{M' k'}-E_{M k}-\hbar sq) \left|M_{M k}^{M'k'}(\bf q)\right|^2 
\nonumber\\
& &\times\biggl(2\Theta(k)\Theta(k')\ T_{M,k} T_{M',k'}
\left\{\rho({\bf q})
\left[ f_{M' k'}\left(1-f_{M' k'}\right)+
f_{M k}\left(1-f_{M k}\right)\right]\right.
\nonumber\\
& &+\left. f_{M' k'}\left(1-f_{M' k'}\right)+
f_{M k}f_{M' k'}\left(f_{M' k'}
-f_{M k}\right)\right\}\nonumber\\
& &+\left[T_{M' k'}-T_{M k}\right]\left\{[\rho({\bf q})+1]
\left[f_{M' k'}\left(1-f_{M' k'}\right)
+f_{M k}f_{M' k'}(f_{M k}+
f_{M' k'}-2)\right]\right.\nonumber\\
& &-\left. \rho({\bf q})\left[f_{M k}\left(1-f_{M k}\right)
+f_{M k}f_{M' k'}(f_{M k}+
f_{M' k'}-2)\right]
\right\}\biggr).
\label{acboltz}
\end{eqnarray}
Here, $M_{M k}^{M' k'}(\bf q)$ is the electron-phonon matrix 
element, 
$f_{M k}\equiv f(E_{M k})$ is the Fermi function with 
temperature $T$ and chemical potential determined by the contact reservoirs, 
and $s$ is the sound velocity. The $x$ coordinate runs along the wire length.   
The only restriction we place on the phonon distribution is 
$\rho(q_x)=\rho(-q_x)$, which means that there is no phonon drag contribution. 
 
The obtained expression consists of two parts. The first part involves only 
the processes of electron backscattering (as can be seen from the 
$\Theta$ stepfunctions) and 
for $\rho({\bf q})=[\exp(\hbar sq/k_{\rm B}T)-1]^{-1}$ 
coincides with the formula given in Ref.\ \cite{GPF95}. We shall call this
the standard part. The second part describes both forward- and backscattering 
processes and is proportional to the difference in transmission probabilities 
for electrons with different energies. We shall call this the
nonstandard part.    

It can   
be easily shown that if $\rho({\bf q})$ is the Bose-Einstein function with the 
same 
temperature as the electron gas, then the nonstandard part disappears. 
Similarly, if the 
transmission probability is energy {\it{in}}dependent (e.g. equal to one), 
then again the nonstandard part vanishes.
This explains why in Refs. \cite{GPF95,BS96,B97} the 
nonstandard
part was absent and AC was noted
as being caused only by backscattering processes.  
Forward phonon scattering will therefore give rise to a nonzero AC if the 
following two conditions are
fulfilled simultaneously:
\begin{itemize}
\item phonons and/or electrons have a nonequilibrium distribution
function and
\item QC has an energy-dependent transmission probability.
\end{itemize}

Let us now discuss some qualitative properties of AC determined by 
general physical 
considerations and independent of the particular potential profile of QC,
details of phonon distribution function and mechanisms of electron-phonon 
interaction. The
resulting picture will be quite rich, since different kinds of electron 
transitions influence AC in  different ways. We consider them in sequence.

As we have already mentioned, the first, standard term in Eq.~(\ref{acboltz}) 
contains only
backscattering processes. They decrease the ballistic current and always cause
negative AC. 
The standard term as a function of gate voltage $V_g$ will have a sharp 
maximum whenever the Fermi level $E_F$ and quantum level
$E_M$ cross \cite{BS96}.

In contrast to the always negative standard term, 
 the nonstandard term in Eq.\ (\ref{acboltz}) containing  
$[T_{M' k'}-T_{M k}]$ may have either sign. Since $T_{M k}$ is a
monotonically increasing function of $E_{M k}$ and $E_{M' k'}>E_{M k}$, so that
$T_{M' k'}-T_{M k}>0$,
this sign is determined only  by the
distribution function part. One can easily show that for
$E_{M k}+E_{M' k'}>2E_F$ the phonon emission part containing 
$[\rho({\bf q})+1]$ is negative whereas the part with $\rho({\bf q})$
describing absorption, is positive. For 
$E_{M k}+E_{M' k'}<2E_F$ the situation is the opposite.
As a result, the sign of $\Delta G$ may depend
on the angle between the current through the QC and the phonon momentum 
${\bf q}$. In the  
experiments of Ref.\ \cite{naylor} 
this angle is close to $\pi/2$ and will be denoted  as  $\pi/2-\theta$. 
It is seen from  
energy and momentum conservation, that a phonon with  given ${\bf q}$ can be
absorbed only by electrons with $k=ms/(\hbar\sin\theta)-(q/2)\sin\theta$ (in 
this
example we consider only intrasubband forward scattering). Thus, for phonons 
with momenta almost normal to the current (small $\theta$), $k$, 
$E_{M k}$
and, hence, $E_{M k}+E_{M' k'}$ will be large and AC positive.
For larger $\theta$, AC becomes negative. 

One further observation can be made concerning AC in the 
`tail' region $E_F<E_1$ (below the onset of
metallic conductivity). In this region electrons in the QC are 
nondegenerate and
their concentration falls rapidly with $E_{M k}$. From the above considerations,  
this means that only phonons with small $\theta$ will be actively 
absorbed and AC caused by the nonstandard term will  always be 
positive. This reflects the evident fact that in the region of activated 
conductivity phonon absorption helps to overcome the barrier 
and increases conductivity.

The  model of a QC considered here has the advantage of simplicity, allowing 
the AC properties to be determined 
without the need for involved numerical
calculations. One  shortcoming of the model, however, is that it
does not account for the fact that in the nonuniform potential of a QC the
electron wavefunctions are not plane waves, which changes the electron-phonon
matrix elements and selection rules. In particular, momentum along the wire
axis is not conserved during an electron-phonon scattering process. This fact 
may be essential in order to explain the large AC signals observed in 
Ref.\ \cite{naylor} where the heater phonon beam was at near normal incidence
to the wire (i.e., phonons would not have enough momentum component $q_x$
to be absorbed if momentum was supposed conserved, and the calculated
AC signal would be much smaller than what was measured \cite{B97}). 

Thus, we must go beyond the simple semiclassical description of AC. An 
alternative approach is to derive a Landauer formula for the QC
conductance which includes the phonon scattering contribution to the electron 
transmission probability through the QC, modeled by some
nonuniform potential \cite{MG97,BS98}.

\acknowledgments

The authors would like to thank the organizers of Phonons 98   
for providing the opportunity to present this work.
Funding by the EPSRC under Grant Nos. GR/K/55493 and GR/L3/722 (visiting
fellowship to A. Y. S.) is also acknowledged.

\end{document}